\documentclass{PoS}%
\usepackage{amsmath}
\usepackage{amsfonts}
\usepackage{amssymb}
\usepackage{graphicx}%
\providecommand{\U}[1]{\protect\rule{.1in}{.1in}}
\title{%
{\protect \vspace{-15mm}
\normalsize \begin{flushright}
KEK Preprint 2015-55 \\
CHIBA-EP-216
\end{flushright}
\vspace{5mm} }
Abelian monopole or non-Abelian monopole responsible for quark confinement%
}

\ShortTitle{Abelian monopole or non-Abelian monopole responsible for quark confinement}
\author{\speaker{Akihiro Shibata}\\
        Computing Research Center, High Energy Accelerator Research Organization (KEK)  \\
and Graduate University for Advanced Studies (Sokendai), Tsukuba 305-0801, Japan\\
        E-mail: \email{Akihiro.Shibata@kek.jp}}

\author{Kei-Ichi Kondo\\
        Department of Physics, Graduate School of Science, Chiba University, Chiba 263-8522, Japan\\
        E-mail: \email{kondok@faculty.chiba-u.jp}}

\author{Seikou Kato\\
       Oyama National College of Technology, Oyama, Tochigi 323-0806, Japan\\
        E-mail: \email{skato@oyama-ct.ac.jp}}

\author{Toru Shinohara\\
        Department of Physics, Graduate School of Science, Chiba University, Chiba 263-8522, Japan\\
        E-mail: \email{sinohara@graduate.chiba-u.jp}}

\abstract{We have pointed out that the $SU(3)$ Yang-Mills theory has a new way of reformulation using new field variables (minimal option), 
in addition to the conventional option adopted by Cho, Faddeev and Niemi (maximal option). The reformulation enables us to change the original
 non-Abelian gauge field into  the new field variables such that one of them called the restricted field gives the dominant contribution 
to quark confinement in the gauge-independent way. In the minimal option, especially, the restricted field is non-Abelian $U(2)$ and involves  
the non-Abelian magnetic monopole. In the preceding lattice conferences, we have accumulated  the numerical evidences 
for the non-Abelian magnetic-monopole dominance in addition to  the restricted non-Abelian field dominance 
for quark confinement supporting the non-Abelian  dual superconductivity using the minimal option for the SU(3) Yang-Mills theory. 
 This should be compared with the maximal option which is a gauge invarient version  of the Abelian projection in the maximal Abelian gauge: 
the restricted field is Abelian  $U(1) \times U(1)$ and involves only the Abelian magnetic monopole, just like the Abelian projection. 

     In this talk, we focus on discriminating between two reformulations, i.e., maximal and minimal options of $SU(3)$ Yang-Mills theory 
for quark confinement from the viewpoint of dual superconductivity. For this purpose, we measure the distribution of the chromoelectric 
flux connecting a quark and an antiquark and the induced magnetic-monopole current around the flux tube.
}

\FullConference{The 33rd International Symposium on Lattice Field Theory\\
		14 -18 July 2015\\
		Kobe International Conference Center, Kobe, Japan*}

\begin{document}
\section{Introduction}

The dual superconductivity is a promising mechanism for quark
confinement.\cite{DualMeisser75}. In order to establish this picture, we have
to show evidences of the dual version of the superconductivity. For this
purpose, we have presented a new formulation of the $SU(3)$ Yang-Mills (YM)
theory using new field variables.(For a review see \cite{PhysRep}) The
reformulation enables us to change the original non-Abelian gauge field into
the new field variables such that one of them called the restricted field
gives the dominant contribution to quark confinement in the gauge-independent
way.\cite{KSM05} The lattice version of a new formulation of $SU(N)$ YM theory
gives the decomposition of a gauge link variable\ corresponding to its
stability gauge group: $U_{x,\mu}=X_{x,\mu}V_{x,\mu}$, where $V_{x,\mu}$ could
be the dominant mode for quark confinement, and $X_{x,\mu}$ the remainder
part.\cite{LatticeYM08}

For $SU(3)$ YM theory, we have two options: the minimal option and maximal
option. In the minimal option, especially, the restricted field is non-Abelian
$U(2)$ and involves the non-Abelian magnetic monopole. In the preceding works,
we have shown numerical evidences of the non-Abelian \ dual superconductivity
using the minimal option for the $SU(3)$ YM theory on a lattice: the
non-Abelian magnetic monopole as well as the restricted non-Abelian field
dominantly reproduces the string tension in the linear potential in $SU(3)$YM
theory \cite{abeliandomSU(3)}, and the $SU(3)$ YM vacuum is the type I dual
superconductor profiled by the chromoelectric flux tube and the magnetic
monopole current induced around it, which is a novel feature obtained by our
simulations. \cite{DMeisner-TypeI2013} We further investigate the
\ confinement/deconfinement phase transition in view of this non-Abelian dual
superconductivity picture.\cite{lattice2013}\cite{lattice2014}\cite{SCGT15}
This should be compared with the maximal option which is adopted first by Cho,
Faddeev and Niemi \cite{CFNS-C}. The restricted field is Abelian $U(1)\times
U(1)$ and involves only the Abelian magnetic monopole.\cite{lattce2007}%
\cite{ChoKundy2014} This is nothing but the gauge invariant version of the
Abelian projection in the maximal Abelian gauge.\cite{Suganuma}%
\cite{suganuma-sakumichi}

In this talk, we focus on discriminating between two reformulations, i.e.,
maximal and minimal options of $SU(3)$YM theory for quark confinement from the
viewpoint of dual superconductivity. For this purpose, we measure string
tension for the restricted non-Abelian field of both minimal and maximal
option in comparison with the string tension for the original YM field. We
also investigate the dual Meissner effect by measuring the distribution of the
chromoelectric flux connecting a quark and an antiquark and the induced
magnetic-monopole current around the flux tube.

\section{Gauge link decompositions}

Let $U_{x,\mu}=X_{x,\mu}V_{x,\mu}$ be a decomposition of the YM link variable
$U_{x,\mu}$, where the YM field and the decomposed new variables are
transformed by the full $SU(3)$ gauge transformation $\Omega_{x}$ such that
$V_{x,\mu}$ is transformed as the gauge link variable and $X_{x,\mu}$ as the
site variable \cite{exactdecomp} :
\begin{subequations}
\label{eq:gaugeTransf}%
\begin{align}
U_{x,\mu}  &  \longrightarrow U_{x,\nu}^{\prime}=\Omega_{x}U_{x,\mu}%
\Omega_{x+\mu}^{\dag},\\
V_{x,\mu}  &  \longrightarrow V_{x,\nu}^{\prime}=\Omega_{x}V_{x,\mu}%
\Omega_{x+\mu}^{\dag},\text{ \ }X_{x,\mu}\longrightarrow X_{x,\nu}^{\prime
}=\Omega_{x}X_{x,\mu}\Omega_{x}^{\dag}.
\end{align}
For the SU(3) YM theory, we have two options discriminated by its stability
group, so called minimal option and maximal option.

\subsection{Minimal option}

The minimal option is obtained for the stability gauge group of $\tilde
{H}=U(2)=SU(2)\times U(1)\subset SU(3)$. By introducing a color field
$\mathbf{h}_{x}=\xi(\lambda^{8}/2)\xi^{\dag}$ $\in\lbrack SU(3)/U(2)]$ with
$\lambda^{8}$ being the Gell-Mann matrix and $\xi$ an $SU(3)$ group element,
the decomposition is given by solving the defining equation:
\end{subequations}
\begin{subequations}
\label{eq:def-min}%
\begin{align}
&  D_{\mu}^{\epsilon}[V]\mathbf{h}_{x}:=\frac{1}{\epsilon}\left[  V_{x,\mu
}\mathbf{h}_{x+\mu}-\mathbf{h}_{x}V_{x,\mu}\right]  =0,\label{eq:def1-min}\\
&  g_{x}:=e^{i2\pi q/3}\exp(-ia_{x}^{0}\mathbf{h}_{x}-i\sum\nolimits_{j=1}%
^{3}a_{x}^{(j)}\mathbf{u}_{x}^{(j)})=1. \label{eq:def2-min}%
\end{align}
Here, the variable $g_{x}$ is an undetermined parameter from
Eq.(\ref{eq:def1-min}), $\mathbf{u}_{x}^{(j)}$ 's are $su(2)$-Lie algebra
valued, and $q$ has an integer value. These defining equations can be solved
exactly \cite{exactdecomp}, and the solution is given by
\end{subequations}
\begin{subequations}
\label{eq:decomp_mini}%
\begin{align}
X_{x,\mu}  &  =\widehat{L}_{x,\mu}^{\dag}\det(\widehat{L}_{x,\mu})^{1/3}%
g_{x}^{-1},\text{ \ \ \ }V_{x,\mu}=X_{x,\mu}^{\dag}U_{x,\mu}=g_{x}\widehat
{L}_{x,\mu}U_{x,\mu},\\
\widehat{L}_{x,\mu}  &  =\left(  L_{x,\mu}L_{x,\mu}^{\dag}\right)
^{-1/2}L_{x,\mu},\\
\text{\ }L_{x,\mu}  &  =\frac{5}{3}\mathbf{1}+\frac{2}{\sqrt{3}}%
(\mathbf{h}_{x}+U_{x,\mu}\mathbf{h}_{x+\mu}U_{x,\mu}^{\dag})+8\mathbf{h}%
_{x}U_{x,\mu}\mathbf{h}_{x+\mu}U_{x,\mu}^{\dag}\text{ .}%
\end{align}
Note that the above defining equations correspond to the continuum version:
$D_{\mu}[\mathcal{V}]\mathbf{h}(x)=0$ and $\mathrm{tr}(\mathbf{h}%
(x)\mathcal{X}_{\mu}(x))$ $=0$, respectively. In the naive continuum limit, we
have reproduced the decomposition $\mathbf{A}_{\mathbf{\mu}}(x)=\mathbf{V}%
_{\mu}(x)+\mathbf{X}_{\mu}(x)$ in the continuum theory \cite{KSM05}
\end{subequations}
\begin{subequations}
\begin{align}
\mathbf{V}_{\mu}(x)  &  =\mathbf{A}_{\mathbf{\mu}}(x)-\frac{4}{3}\left[
\mathbf{h}(x),\left[  \mathbf{h}(x),\mathbf{A}_{\mathbf{\mu}}(x)\right]
\right]  -ig^{-1}\frac{4}{3}\left[  \partial_{\mu}\mathbf{h}(x),\mathbf{h}%
(x)\right]  ,\\
\mathbf{X}_{\mu}(x)  &  =\frac{4}{3}\left[  \mathbf{h}(x),\left[
\mathbf{h}(x),\mathbf{A}_{\mathbf{\mu}}(x)\right]  \right]  +ig^{-1}\frac
{4}{3}\left[  \partial_{\mu}\mathbf{h}(x),\mathbf{h}(x)\right]  .
\end{align}

The decomposition (\ref{eq:decomp_mini}) is uniquely obtained, if color
fields$\{\mathbf{h}_{x}\}$ are obtained. To determine the configuration of
color fields, we use the reduction condition to formulate the new theory
written by new variables ($X_{x,\mu}$,$V_{x,\mu}$) which is equipollent to the
original YM theory. Here, we use the reduction functional:
\end{subequations}
\begin{equation}
F_{\text{red}}[\mathbf{h}_{x}]=\sum\nolimits_{x,\mu}\mathrm{tr}\left\{
(D_{\mu}^{\epsilon}[U_{x,\mu}]\mathbf{h}_{x})^{\dag}(D_{\mu}^{\epsilon
}[U_{x,\mu}]\mathbf{h}_{x})\right\}  ,\label{eq:reduction-min}%
\end{equation}
and then color fields $\left\{  \mathbf{h}_{x}\right\}  $ are obtained by
minimizing the functional (\ref{eq:reduction-min}).

\subsection{Maximal option}

The maximal option is obtained for the stability group of the maximal torus
group $\tilde{H}=U(1)\times U(1)\subset SU(3)$. By introducing a set of color
fields $\mathbf{n}^{(3)}=\xi(\lambda^{3}/2)\xi^{\dag},\ \mathbf{n}^{(8)}%
=\xi(\lambda^{8}/2)\xi^{\dag}$ $\in\lbrack SU(3)/U(2)]$ with $\lambda
^{3},\lambda^{8}$ being the Gell-Mann matrices and $\xi$ an $SU(3)$ group
element, \ the decomposition is given by solving the defining equation:
\begin{subequations}
\label{eq:DefEq}%
\begin{align}
&  D_{\mu}^{\epsilon}[V]\mathbf{n}_{x}^{(j)}:=\frac{1}{\epsilon}\left[
V_{x,\mu}\mathbf{n}_{x+\mu}^{(j)}-\mathbf{n}_{x}^{(j)}V_{x,\mu}\right]
=0\text{ \ \ \ }j=3,8,\label{eq:def1_max}\\
&  g_{x}:=e^{i2\pi q/3}\exp(-ia_{x}^{3}\mathbf{n}_{x}^{(3)}-ia_{x}%
^{(8)}\mathbf{n}_{x}^{(8)})=1.\label{eq:def2_max}%
\end{align}
The variable $g_{x}$ is an undetermined parameter from Eq.(\ref{eq:def1_max})
and $q$ has an integer value. These defining equations can be solved exactly
\cite{exactdecomp}, and the solution is given by
\end{subequations}
\begin{subequations}
\label{eq:decomp_max}%
\begin{align}
X_{x,\mu} &  =\widehat{K}_{x,\mu}^{\dag}\det(\widehat{K}_{x,\mu})^{1/3}%
g_{x}^{-1},\text{ \ \ \ }V_{x,\mu}=X_{x,\mu}^{\dag}U_{x,\mu}=g_{x}\widehat
{K}_{x,\mu}U_{x,\mu},\\
\widehat{K}_{x,\mu} &  =\left(  K_{x,\mu}K_{x,\mu}^{\dag}\right)
^{-1/2}K_{x,\mu},\text{ \ \ \ \ \ \ }\\
K_{x,\mu} &  =\mathbf{1}+6(\mathbf{n}_{x}^{(3)}U_{x,\mu}\mathbf{n}_{x+\mu
}^{(3)}U_{x,\mu}^{\dag})+6(\mathbf{n}_{x}^{(8)}U_{x,\mu}\mathbf{n}_{x+\mu
}^{(8)}U_{x,\mu}^{\dag})
\end{align}
Note that the above defining equations correspond to the continuum version:
$D_{\mu}[\mathcal{V}]\mathbf{n}^{(j)}(x)=0$ and $\mathrm{tr}(\mathbf{n}%
^{(j)}(x)\mathcal{X}_{\mu}(x))$ $=0$, respectively. In the naive continuum
limit, we have reproduced the decomposition $\mathbf{A}_{\mathbf{\mu}%
}(x)=\mathbf{V}_{\mu}(x)+\mathbf{X}_{\mu}(x)$ in the continuum theory as
\cite{KSM05}\cite{CFNS-C}
\end{subequations}
\begin{subequations}
\begin{align}
\mathbf{V}_{\mu}(x) &  =\sum\nolimits_{j=3,8}\left\{  2\mathrm{tr}\left(
\mathbf{A}_{\mathbf{\mu}}(x)\mathbf{n}^{(j)}(x)\right)  \mathbf{n}%
^{(j)}(x)-ig^{-1}\left[  \partial_{\mu}\mathbf{n}^{(j)}(x),\mathbf{n}%
^{(j)}(x)\right]  \right\}  ,\\
\mathbf{X}_{\mu}(x) &  =\sum\nolimits_{j=3,8}\left[  \mathbf{n}^{(j)}%
(x),\left[  \mathbf{n}^{(j)}(x),\mathbf{A}_{\mathbf{\mu}}(x)\right]  \right]
.
\end{align}
To determine the configuration of color fields, we use the reduction condition
to formulate the new theory written by new variables ($X_{x,\mu}$,$V_{x,\mu}$)
which is equipollent to the original YM theory. Here, we use the reduction
functional:
\end{subequations}
\begin{equation}
F_{\text{red}}[\mathbf{n}_{x}^{(3)},\mathbf{n}_{x}^{(8)}]=\sum\nolimits_{x,\mu
}\sum\nolimits_{j=3,8}\mathrm{tr}\left\{  (D_{\mu}^{\epsilon}[U_{x,\mu
}]\mathbf{n}_{x}^{(j)})^{\dag}(D_{\mu}^{\epsilon}[U_{x,\mu}]\mathbf{n}%
_{x}^{(j)})\right\}  ,\label{eq:reduction-max}%
\end{equation}
and then color fields $\left\{  \mathbf{n}_{x}^{(3)},\mathbf{n}_{x}%
^{(8)}\right\}  $ are obtained by minimizing the functional
(\ref{eq:reduction-max}).

It should be noticed that the the resulting decomposition is the gauge
invariant version of the Abelian projection in the maximal Abelian (MA)
gauge.  Because by using the gauge transformation ${}^{G}U_{x,\mu}=\xi
_{x}^{\dag}U_{x,\mu}\xi_{x+\mu}$, the reduction functional
(\ref{eq:reduction-max}) is rewritten into the functional for the MA  gauge
fixing:
\begin{equation}
F_{\text{red}}=\sum\nolimits_{x,\mu}\left\{  1-\frac{1}{4}\mathrm{tr}\left(
{}^{G}U_{x,\mu}\lambda_{3}{}^{G}U_{x,\mu}^{\dag}\lambda_{3}\right)  +\frac
{1}{4}\mathrm{tr}\left(  {}^{G}U_{x,\mu}\lambda_{8}{}^{G}U_{x,\mu}^{\dag
}\lambda_{8}\right)  \right\}
\end{equation}
Then, we can show that the decomposition of the $V$-field in
Eq.(\ref{eq:decomp_max}) is rewritten into the diagonal part of the YM field
in the MA gauge, i.e., Abelian projection in the MA gauge:%
\begin{equation}
{}^{(MAG)}V_{x,\mu}=\mathrm{diag}\left(  \left.  \left(  ^{G}U_{x,\mu}\right)
_{11}\right/  \left\vert \left(  ^{G}U_{x,\mu}\right)  _{11}\right\vert
,\left.  \left(  ^{G}U_{x,\mu}\right)  _{22}\right/  \left\vert \left(
^{G}U_{x,\mu}\right)  _{22}\right\vert ,\left.  \left(  ^{G}U_{x,\mu}\right)
_{33}\right/  \left\vert \left(  ^{G}U_{x,\mu}\right)  _{33}\right\vert
\right)  .
\end{equation}

\section{Lattice result}

We generate the YM gauge field configurations (link variables) $\{U_{x,\mu}\}$
using the standard Wilson action. We prepare 500 data sets on the lattice with
the size of $24^{4}$ at $\beta=2N_{c}/g^{2}\ (N_{c}=3)$: $\beta=6.2$ every 800
sweeps after 10000 thermalization. We obtain two types of decomposed gauge
link variables $U_{x,\mu}=X_{x,\mu}V_{x,\mu}$ for each gauge link by using the
formula Eqs.(\ref{eq:decomp_mini}) and (\ref{eq:decomp_max}) given in the
previous section, after the color-field configuration $\{\mathbf{h}_{x}\}$ and
$\left\{  \mathbf{n}_{x}^{(3)},\mathbf{n}_{x}^{(8)}\right\}  $ are obtained by
solving the reduction condition of the functional (\ref{eq:reduction-min}) and
(\ref{eq:reduction-max}), respectively. In the measurement of the Wilson loop
average defined below, we apply the APE smearing technique to reduce noises.

\subsection{Static potential}

\begin{figure}[ptb]
\begin{center}
\includegraphics[height=4.6cm, angle=270]
{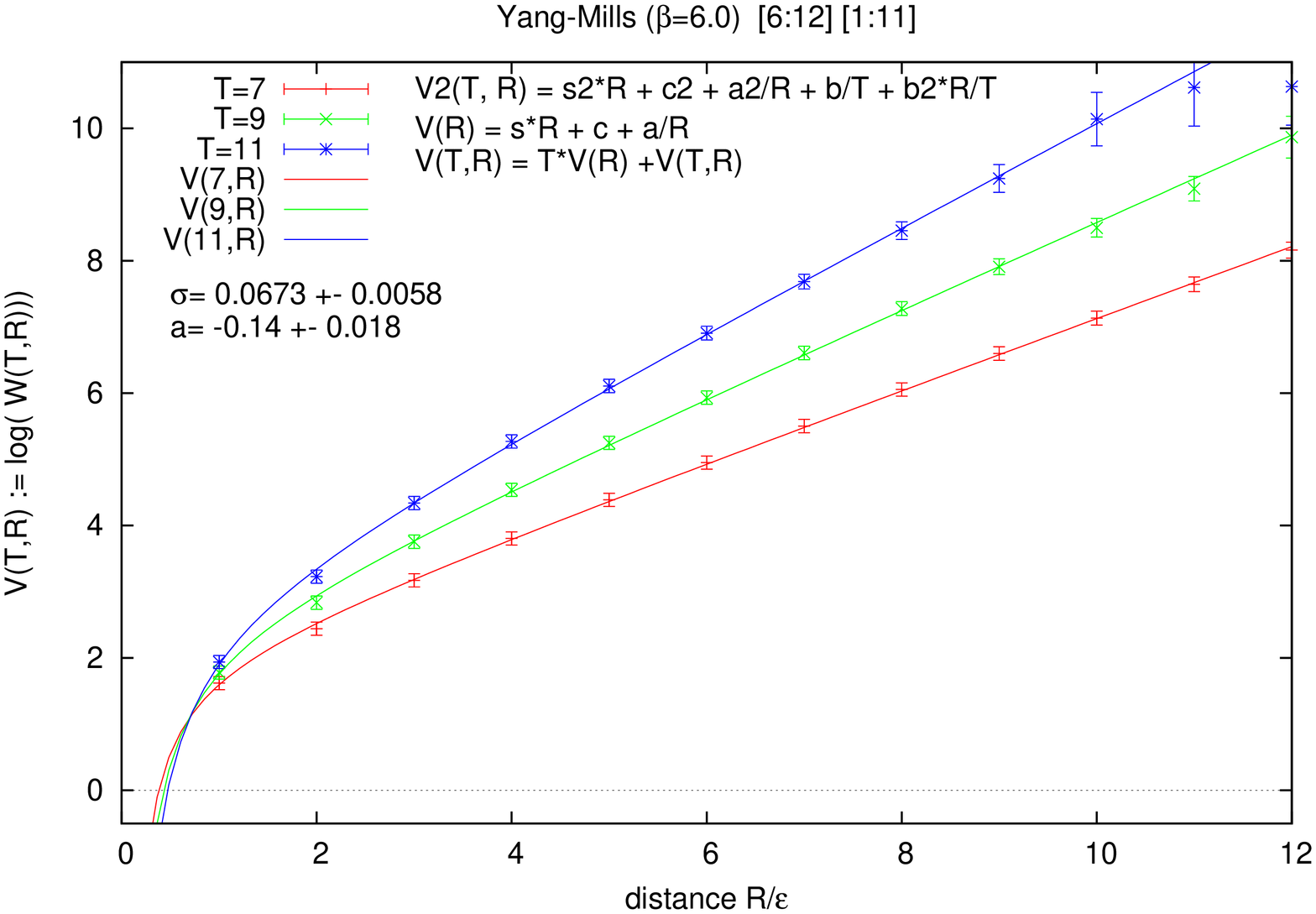} \includegraphics[height=4.6cm, angle=270]
{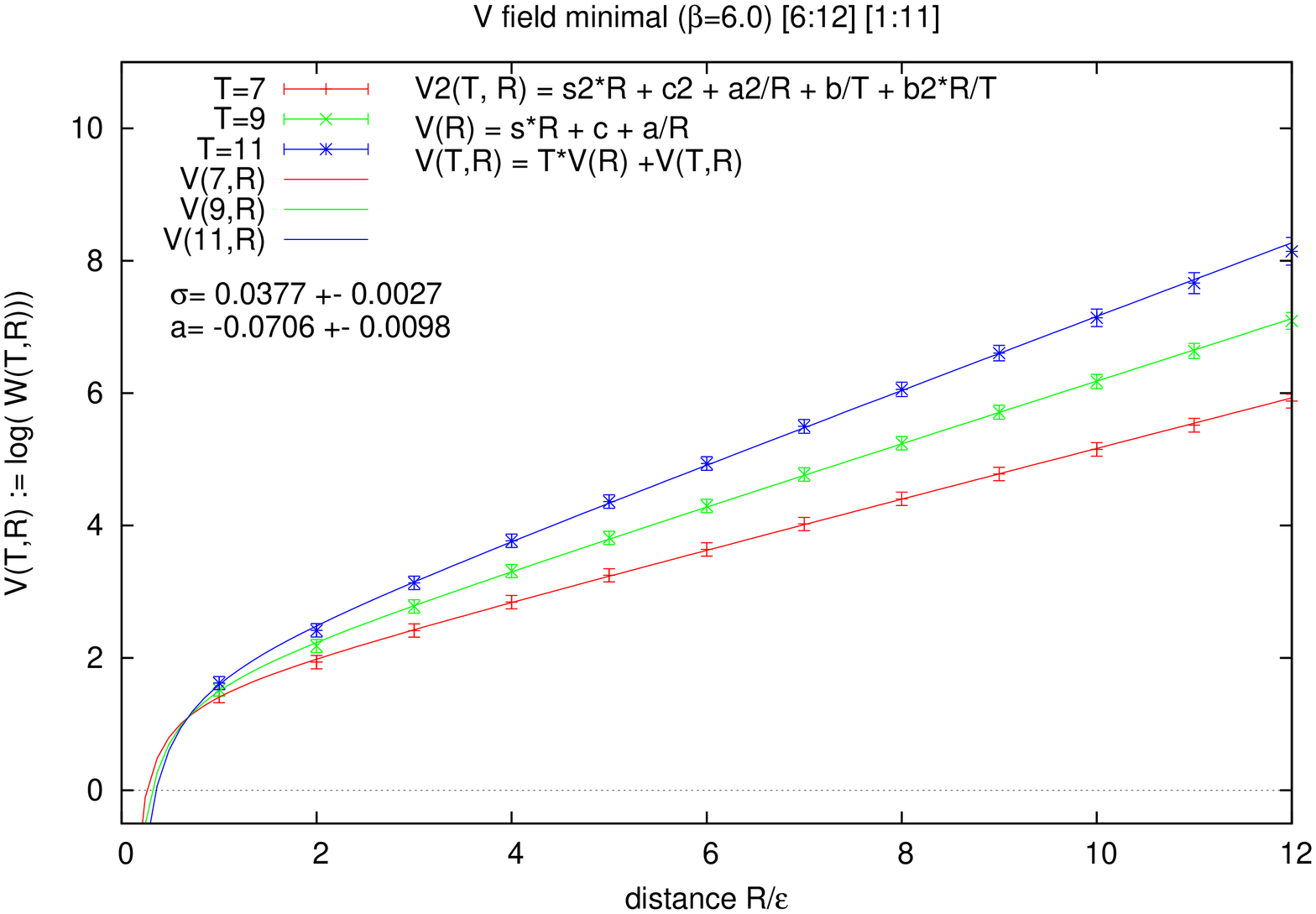} \includegraphics[height=4.6cm, angle=270]
{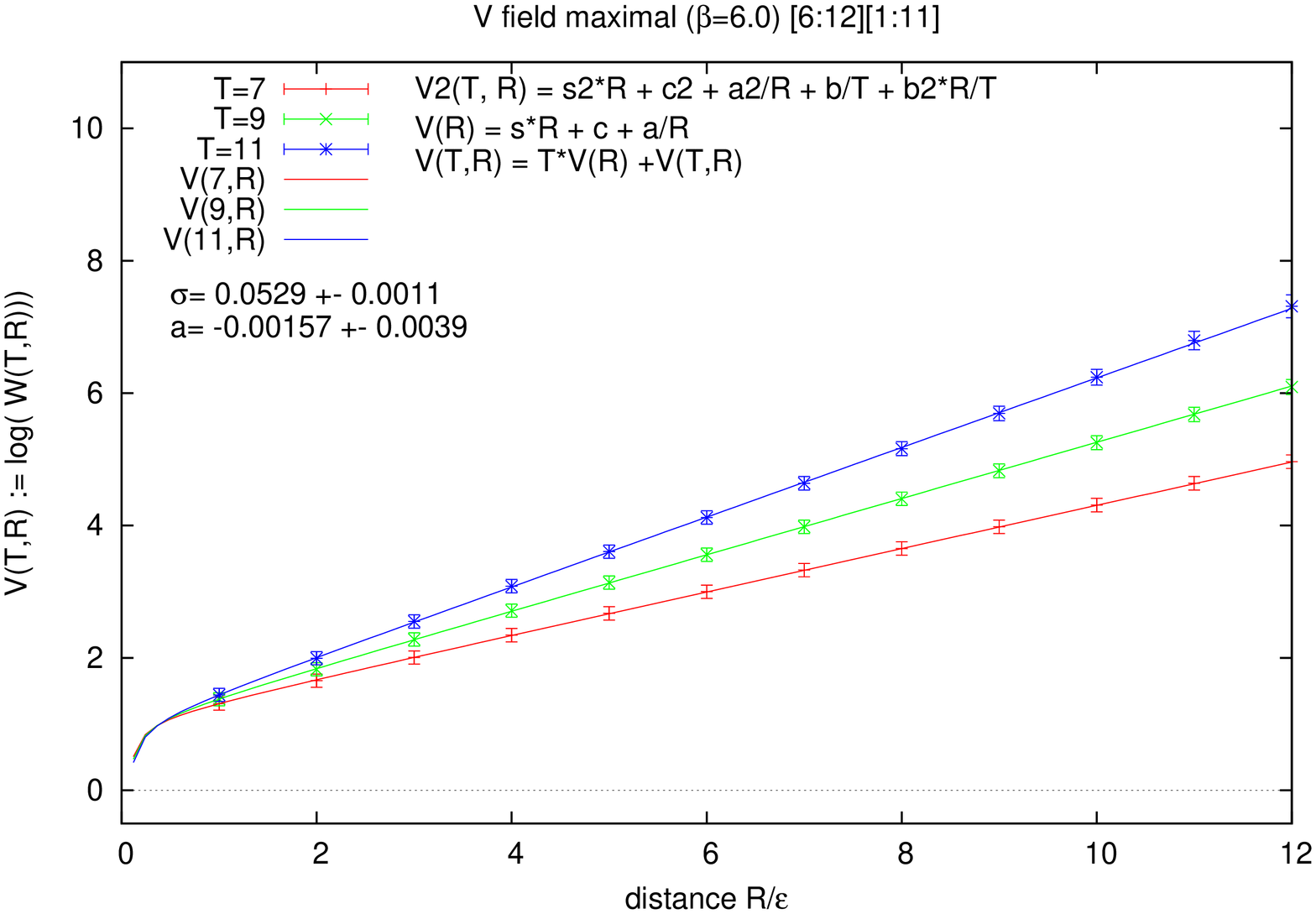}
\end{center}
\caption{Static potential as a function of $R$: Panels show static potentials
for YM field (left), for restricted field in the minimal option (center) and
for restriced field in the maximal option (right). Each panel shows plots of
data for $N_{T}=7,9,11$ cases, and fitted functions $V(T,R)$. }%
\label{fig:potential}%
\end{figure}

We\ first study the static potential from the Wilson loop $C$ of the $T\times
R$ rectangle for both restricted fields of minimal option and maximal option
in addition to the original YM\ field:%
\begin{equation}
W_{\min}(T,R):=\prod\nolimits_{<x,\mu>\in C}V_{x,\mu}^{\min}\text{,
\ \ }W_{\max}(T,R):=\prod\nolimits_{<x,\mu>\in C}V_{x,\mu}^{\max}\text{,
\ \ }W_{\text{YM}}(T,R):=\prod\nolimits_{<x,\mu>\in C}U_{x,\mu}\text{ ,}%
\end{equation}
respectively. To obtain the static potential, we apply a fitting formula:
\begin{align}
&  -\log\left\langle W(T,R)\right\rangle =V(T,R)=T\times V(R)\ +V_{2}(T,R)\\
&  V(R)=\sigma R+c+a/R\text{, \ }V_{2}(T,R)=\sigma_{2}R+c_{2}+a_{2}%
/R+b/T+b_{2}R/T
\end{align}
to data of Wilson loop average $\left\langle W(T,R)\right\rangle $ with
$R=1,2,\cdots,11$ and $T=7,\cdots,11.$ The conventional Cornell potential is
given by the $V(R)$ part.\ \ Figure \ref{fig:potential} shows the fitting
results. Panels from left to right show data and fitting with $N_{T}=7,8,9$
for the original YM field and the restricted fields of minimal and maximal
options, respectively. We find the restricted field dominance in the string
tension for both minimal and maximal options.

\begin{figure}[tbp] \centering
\begin{minipage}{0.38\textwidth} \centering
\includegraphics[
height=5.5cm, angle=270]
{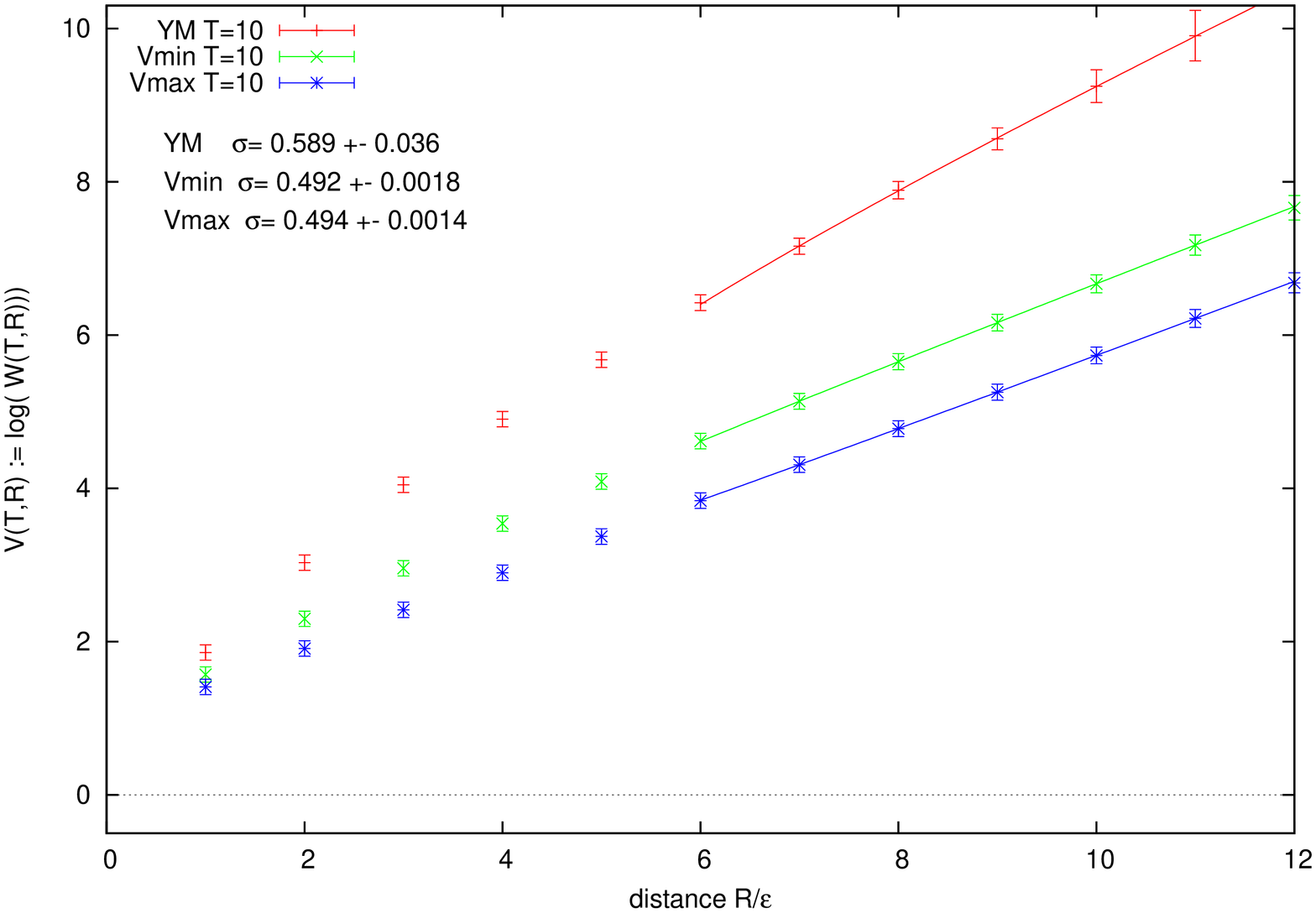}%
\caption{Comparison of static potentials:.  the YM filed,  the restricted field
in the minimal and maximal options. }\label{fig:potential-comp}%
\end{minipage}
\ \ \
\begin{minipage}{0.58\textwidth} \centering
\includegraphics[width=40mm]
{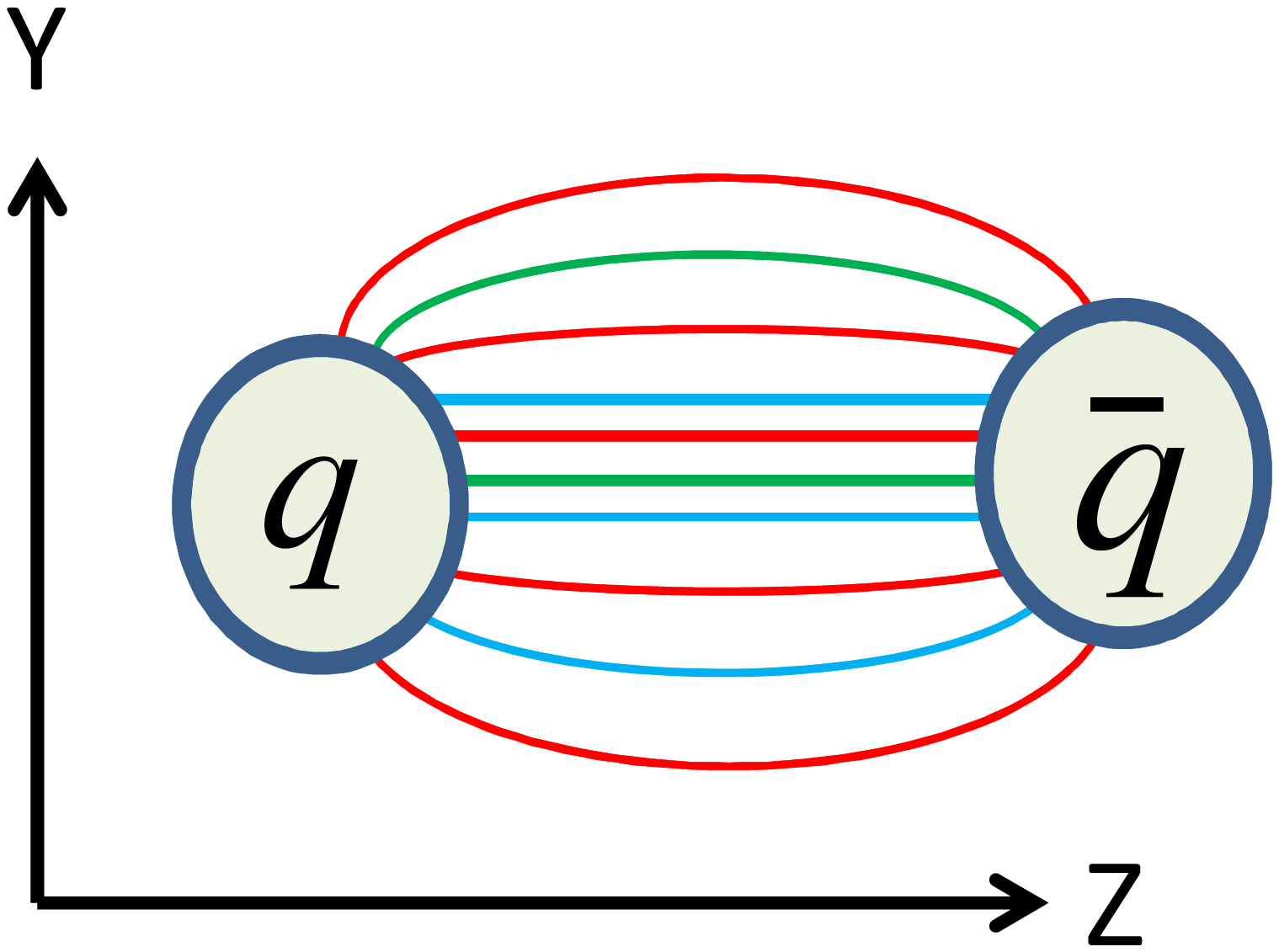}
\ \includegraphics[width=40mm]
{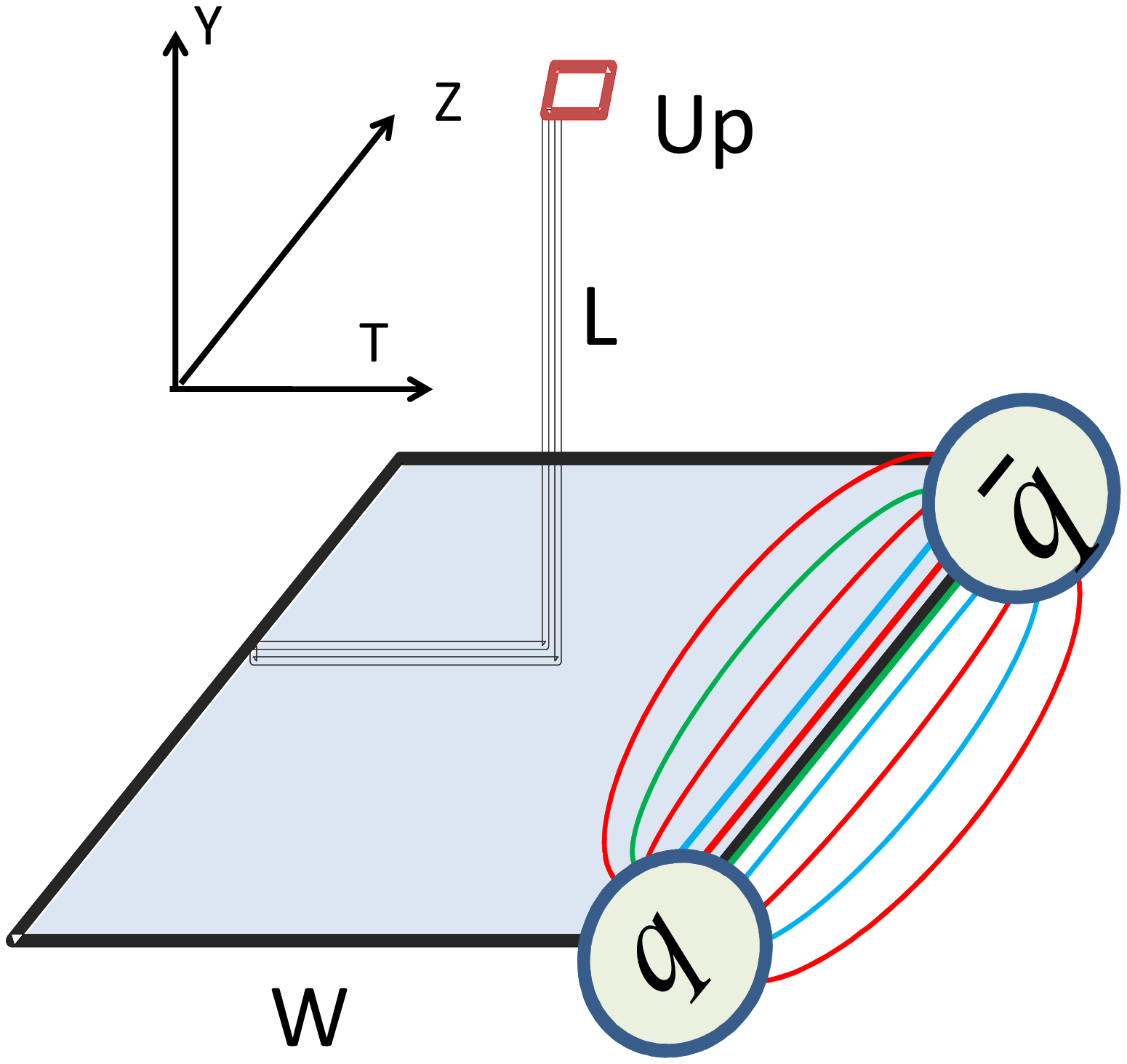}%
\caption{ (Left) Set up of mesuarement of chromo flux.
(Right) The gauge-invariant correlation operator $tr(WLU_{p}L^{\dagger
})$ between a probe plaquatte  $U_{p}$
and the source Wilson-loop $W$. .}%
\label{fig:measure-flux}%
\end{minipage}
\end{figure}

Figure \ref{fig:potential-comp} shows the combined plot of
potentials for $N_{T}=10.$ We obtain string tension in the fitting range
$6\leq R\leq11$ as
\begin{equation}
\sigma_{YM}=0.589\pm0.036\text{, \ \ \ }\sigma_{\min}=0.492\pm0.0018,\text{
\ }\sigma_{\max}=0.494\pm0.0014\text{ \ .}%
\end{equation}
We have\ the good agreement in the string tension between both options.%

\subsection{Chromoelectric Flux}

Next, we study the dual Meissner effect. For this purpose, we measure the
chromo flux created by a quark-antiquark pair which is represented by the
Wilson loop $W$ defined in the right panel of Figure \ref{fig:measure-flux}.
The chromo-field strength, i.e., the field strength of the chromo flux at the
position $P$ is measured by using a plaquette variable $U_{p}$ as the probe
operator for the field strength \cite{Giacomo}:
\begin{equation}
\rho_{_{U_{P}}}:=\frac{\left\langle \mathrm{tr}\left(  WLU_{p}L^{\dag}\right)
\right\rangle }{\left\langle \mathrm{tr}\left(  W\right)  \right\rangle
}-\frac{1}{N_{c}}\frac{\left\langle \mathrm{tr}\left(  U_{p}\right)
\mathrm{tr}\left(  W\right)  \right\rangle }{\left\langle \mathrm{tr}\left(
W\right)  \right\rangle },\label{eq:Op}%
\end{equation}
where $L$ is the Wilson line connecting the source $W$ and the probe $U_{p}$
needed to obtain the gauge-invariant result. Note that $\rho_{_{U_{P}}}$ is
sensitive to the field strength rather than the disconnected one. To
discriminate the chromo flux for each option, for the same source represented
by the Wilson-loop made of the YM field we investigate the chromo flux by
changing probe operators, $LU_{p}L^{\dag},$made of the restricted fields in
the minimal and maximal options in place of the original YM\ field.

\begin{figure}[ptb]
\begin{center}
\includegraphics[
height=5cm, angle=270]
{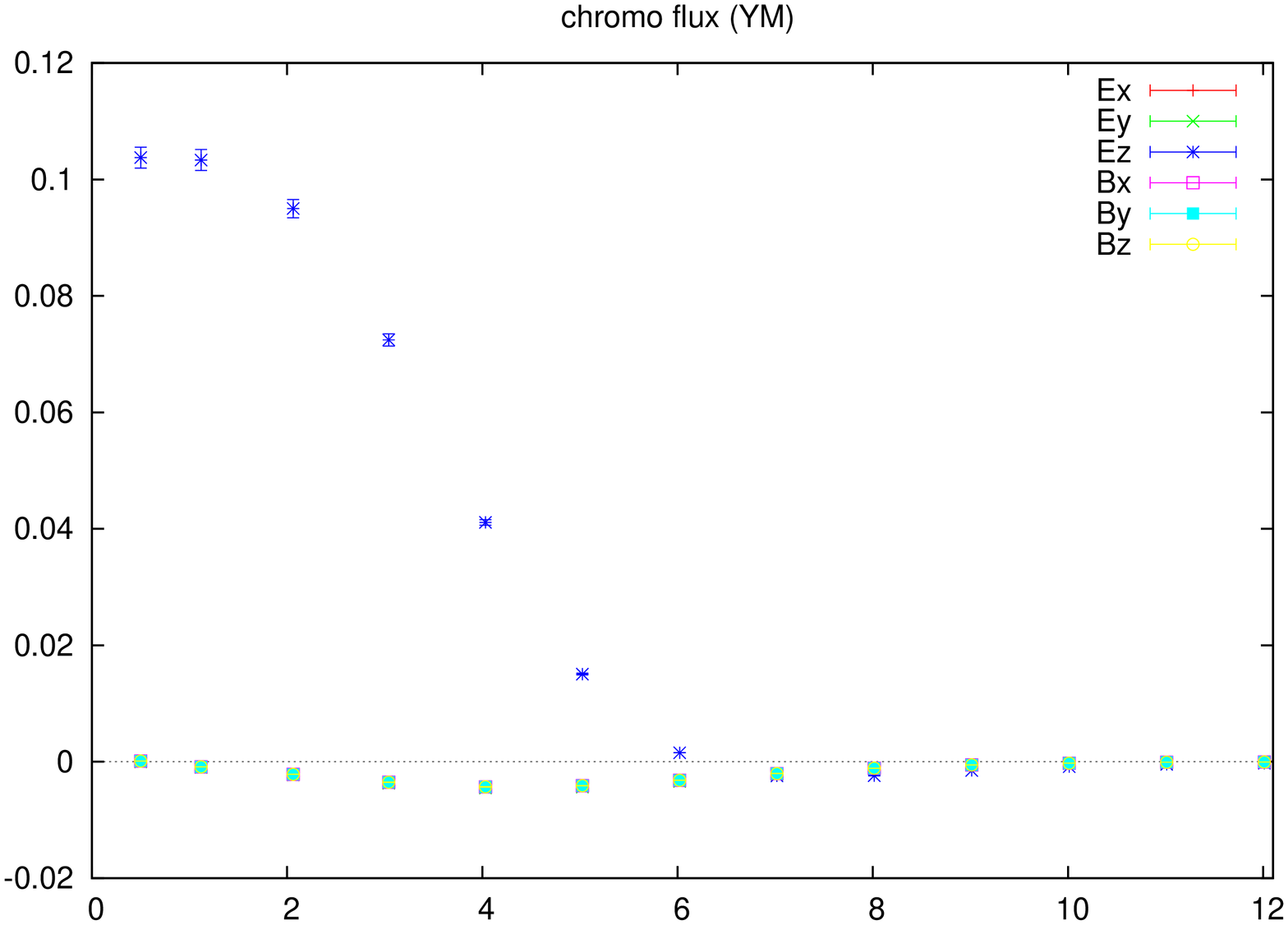}\includegraphics[
height=5cm, angle=270]
{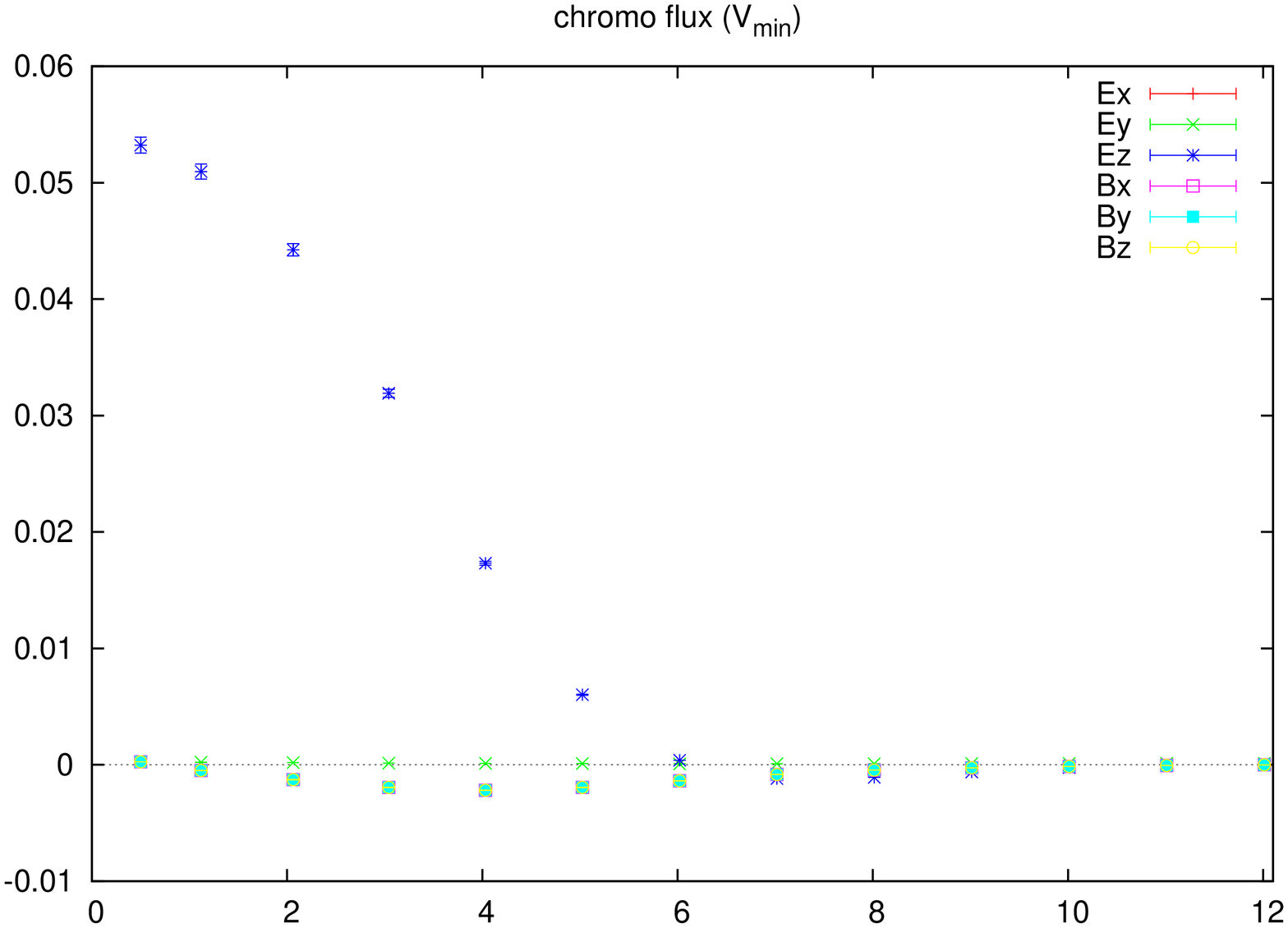}\includegraphics[
height=5cm, angle=270]
{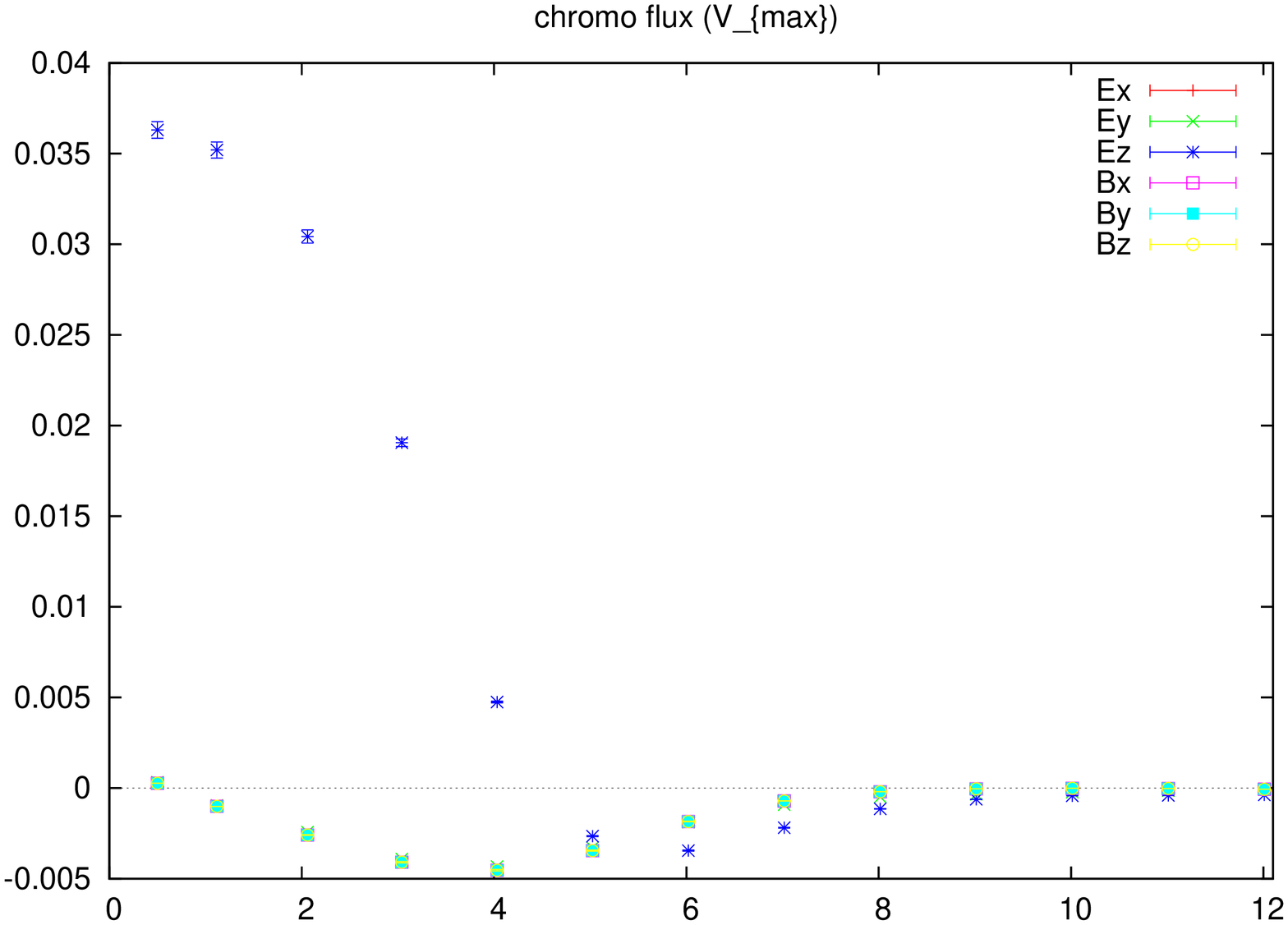}
\end{center}
\caption{Measurment of the componet of the chromo flux. (Left) original
YM\ field, (Center) ninimal option (Right) maximal option.}%
\label{fig:chromo-flux}%
\end{figure}

Figure \ref{fig:chromo-flux} shows the result of the measurement of chromo
field strength for the original YM field, restricted field of the minimal and
maximal options from left to right panels, respectively. We find that only the
chromoelectric-flux tube is created between a quark and an anti-quark for both options.

\begin{figure}[ptbh]
\begin{center}
\includegraphics[width=4.0cm, origin=c]{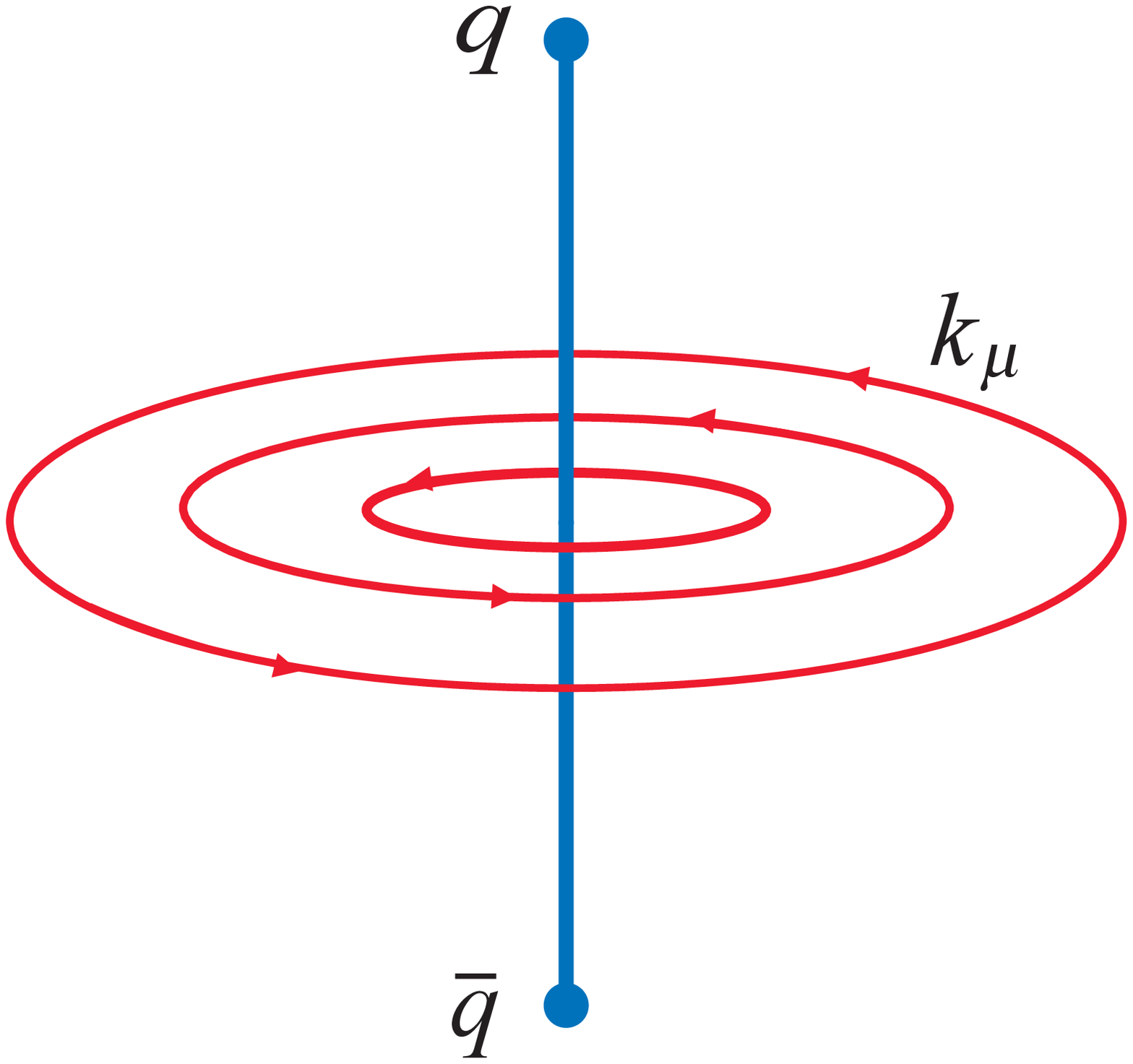}
\includegraphics[ origin=c,
height=5cm, angle=270]
{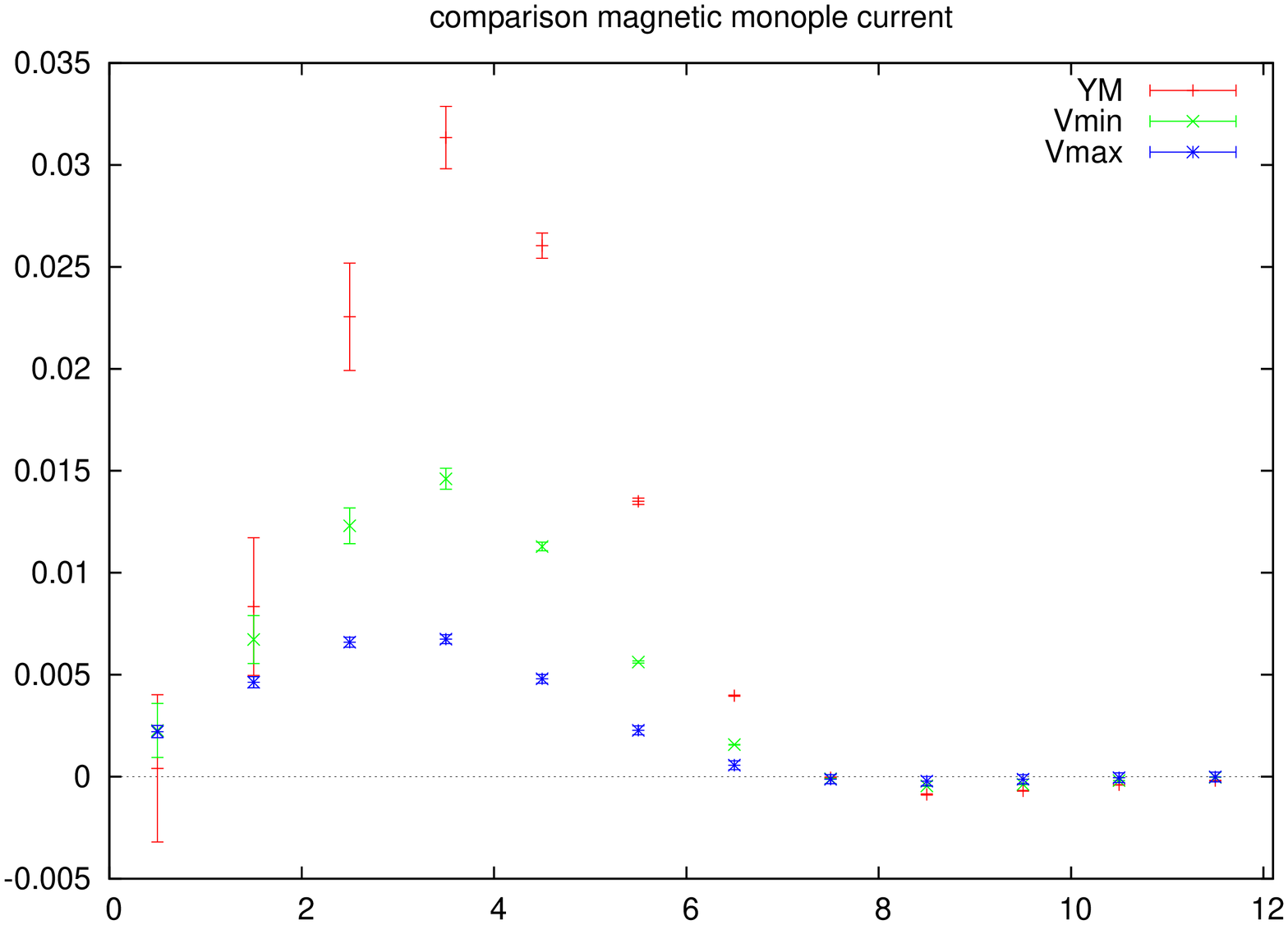}
\includegraphics[ origin=c,
height=5cm, angle=270]
{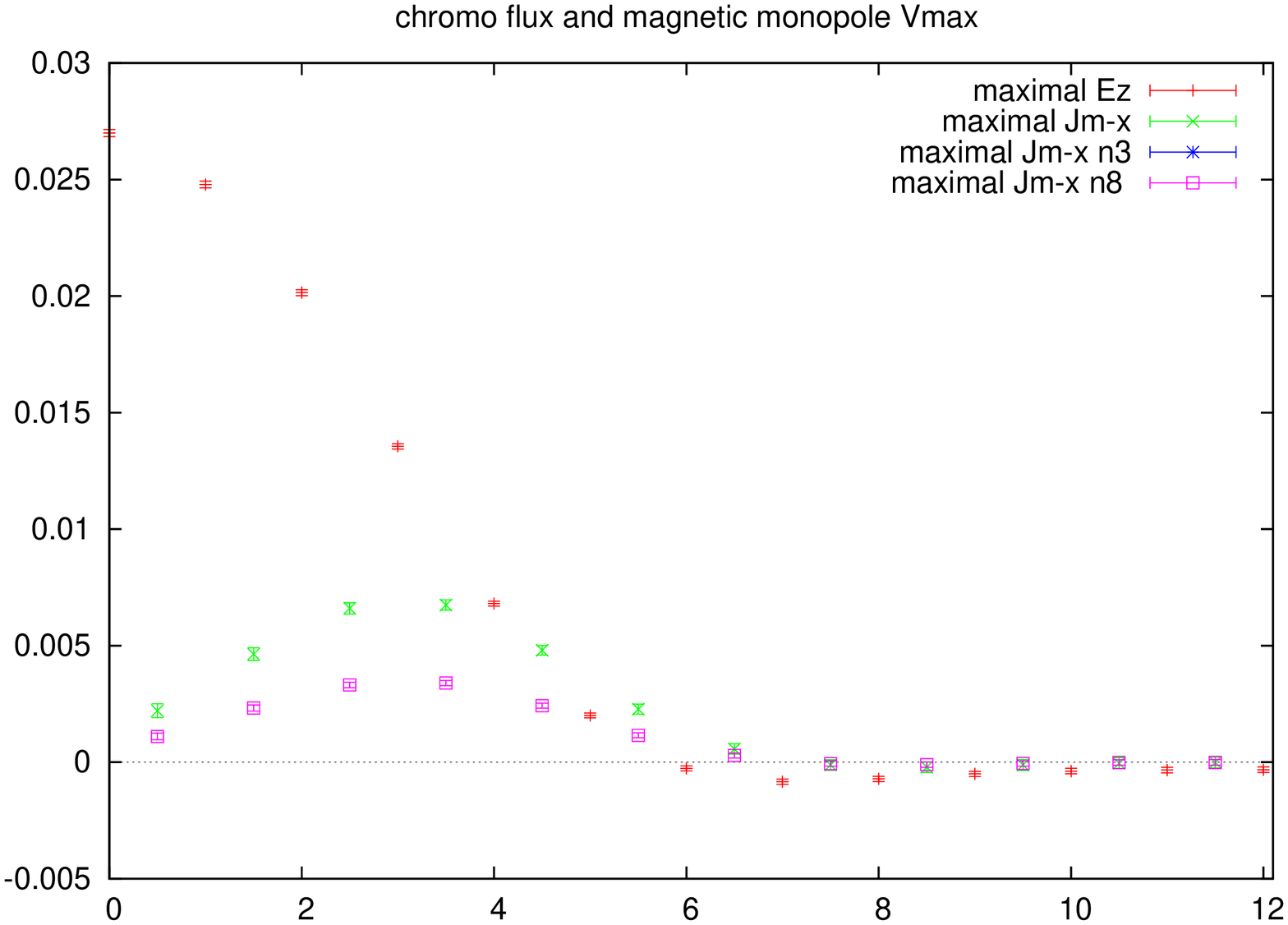} \vspace{-10mm}
\end{center}
\caption{Measurement of the induced magnetic-monopole current. (Left) Sketch
of induced magnetic monopole current around the chromoelectric flux tube.
(Center) Measurement of the induced magetic monopole currents for YM-field
(red), minimal option (green) and maximal option (blue). (Right) Anatomy of
the induced magnetic-monopole current for the maximal option. The upper to
lower plots represent respectively the chromoelectric field ($E_{Z})$ and the
total induced magnetic-monopole current, and decomposed ones corresponding to
the color field $n^{(3)}$ and $n^{(8)}$. }%
\label{fig:magnetic-monopole}%
\end{figure}

Finally, we investigate the dual Meissner effect by measuring the
magnetic--monopole current $k$ induced around the chromoelectric-flux tube
created by the quark-antiquark pair. We use the magnetic-monopole current $k$
defined by
\begin{equation}
k_{\mu}(x)=\frac{1}{2}\epsilon_{\mu\nu\alpha\beta}\left(  F[V]_{\alpha\beta
}(x+\hat{\nu})-F[V]_{\alpha\beta}(x)\right)  . \label{eq:m_current}%
\end{equation}
Note that the magnetic--monopole current (\ref{eq:m_current}) must vanish due
to the Bianchi identity, if there exists no singularity in the gauge
potential. Therefore, the magnetic--monopole current defined in this way can
be the order parameter for the confinement/deconfinement phase transition, as
suggested from the dual superconductivity hypothesis (see left panel of Figure
\ref{fig:magnetic-monopole}). Center panel of Fig.\ref{fig:magnetic-monopole}
shows the result of the measurements of the magnitude $\sqrt{k_{\mu}k_{\mu}}$
of the induced magnetic current $k_{\mu}$ obtained according to
Eq.(\ref{eq:m_current}). Therefore, we find the dual Meissner effect in both
options. Furthermore, in the maximal option we can decompose the
magnetic-monople current into the two channels, $k_{\mu}(x)=$ $k_{\mu}%
^{(3)}(x)+k_{\mu}^{(8)}(x)$ which correspond to two types of color fields
$\mathbf{n}_{x}^{(3)}$ and $\mathbf{n}_{x}^{(8),}$respectively. Right panel of
Fig.\ref{fig:magnetic-monopole} shows the result of the decomposition. This
shows that two types of magnetic monopole contribute to confinement in the
same weight.

\section{Summary and outlook}

By using a new formulation of YM theory, we have investigated two possible
types of the dual superconductivity picture in the $SU(3)$ YM theory, i.e.,
the non-Abelian dual superconductivity as the minimal option and the
conventional Abelian dual superconductivity as the maximal option. In the
measurement for both maximal and minimal options, we have found the linear
static potential between a pair of quark and antiquark as for the original
YM\ field and also the V-field dominance in the string tension for each
option. The string tension for both options has almost the same value \ We,
then, have investigated the dual Meissner effect and found the
chromoelectoric-flux tube in each option. We have also found the induced
magnetic-monopole current due to the dual Meissner effect.

\subsection*{Acknowledgement}

This work is supported by Grant-in-Aid for Scientific Research (C) 24540252
and 15K05042 from Japan Society for the Promotion Science (JSPS), and in part
by JSPS Grant-in-Aid for Scientific Research (S) 22224003. The numerical
calculations are supported by the Large Scale Simulation Program No.13/14-23
(2013-2014) and No.14/15-24 (2014-2015) of High Energy Accelerator Research
Organization (KEK).


\begin{thebibliography}{99}                                                                                               %


\bibitem {PhysRep}Kei-Ichi Kondo, Seikou Kato, Akihiro Shibata and Toru
Shinohara, Phys.Rept. 579 (2015) 1-226, arXiv:1409.1599 [hep-th]

\bibitem {DualMeisser75}Y. Nambu, Phys. Rev. D10, 4262(1974); G. 't Hooft, in
High Energy Physics, edited by A. Zichichi (Editorice Compositori, Bologna,
1975); S. Mandelstam, Phys. Report 23, 245(1976); A.M. Polyakov, Nucl. Phys.
B120, 429(1977).

\bibitem {KSM05}K.-I. Kondo, T. Murakami and T. Shinohara, Eur. Phys. J. C 42,
475 (2005); K.-I. Kondo, T. Murakami and T. Shinohara, Prog. Theor. Phys. 115,
201 (2006).; K.-I. Kondo, T. Shinohara and T. Murakami, Prog.Theor. Phys. 120,
1 (2008)

\bibitem {LatticeYM08}K.-I. Kondo, A. Shibata, S. Kato, T. Shinohara, T.
Murakami, Phys.Lett. B669 (2008) 107-118

\bibitem {abeliandomSU(3)}K.-I. Kondo, A. Shibata, T. Shinohara, S. Kato,
Phys.Rev. D83 (2011) 114016

\bibitem {DMeisner-TypeI2013}A. Shibata, K.-I. Kondo, S. Kato and T.
Shinohara, Phys.Rev. D87 (2013) 5, 05401

\bibitem {lattice2013}Akihiro Shibata, Kei-Ichi Kondo, Seikou Kato, Toru
Shinohara, PoS LATTICE2013 (2014)

\bibitem {lattice2014}A. Shibata, K.-I. Kondo, S. Kato and T. Shinohara, PoS
LATTICE2014 (2015) 340 KEK-PREPRINT-2014-43, CHIBA-EP-210

\bibitem {SCGT15}A. Shibata, K.-I. Kondo, S. Kato and T.Shinohara,
KEK-PREPRINT-2015-49 CHIBA-EP-214

\bibitem {exactdecomp}A. Shibata, K.-I. Kondo and T. Shinohara,
Phys.Lett.B691:91-98 (2010)

\bibitem {CFNS-C}Y.M. Cho, Phys. Rev. D 21, 1080 (1980). Phys. Rev. D 23, 245
(1981); Y.S. Duan and M.L. Ge, Sinica Sci., 11, 1072(1979); L. Faddeev and
A.J. Niemi, Phys. Rev. Lett. 82, 1624 (1999); S.V. Shabanov, Phys. Lett. B
458, 322 (1999). Phys. Lett. B 463, 263 (1999).

\bibitem {lattce2007}A. Shibata, S. Kato, K.-I. Kondo, T. Shinohara \ and S.
Ito, POS(LATTICE2007) 331

\bibitem {ChoKundy2014}Nigel Cundy, Y.M. Cho, Weonjong Lee, Jaehoon Leem,
Phys.Lett. B729 192-198 (2014)

\bibitem {Suganuma}Shinya Gongyo, Takumi Iritani and Hideo Suganuma, Phys.Rev.
D86 (2012) 094018

\bibitem {suganuma-sakumichi}H. Suganuma and N. Sakumichi, Phys.Rev. D90
(2014) 11, 111501

\bibitem {Giacomo}A. Di Giacomo, M. Maggiore, and S. Olejnik, Phys. Lett.
B236, 199 (1990); Nucl. Phys. B347, 441 (1990).
\end{thebibliography}
\end{document}